\documentclass[twocolumn,amsmath,preprintnumbers,amssymb,superscriptaddress,nofootinbib]{revtex4}
\usepackage{hyperref}
\usepackage{graphicx}
\newcommand {\beq}{\begin{equation}}
\newcommand {\eeq}{\end{equation}}
\newcommand {\beqa}{\begin{eqnarray}}
\newcommand {\eeqa}{\end{eqnarray}}
\newcommand {\n}{\nonumber \\}

\newcommand {\p}{\partial}
\newcommand {\tr}{{\rm tr}}
\newcommand {\Tr}{{\rm Tr}}

\begin{document}

\preprint{KEK-TH-1168}

\title{Green-Schwarz superstring from type IIB matrix model}

\date{published 28 January 2008}

\author{Yoshihisa Kitazawa}
\email{kitazawa@post.kek.jp}
\affiliation{High Energy Accelerator Research Organization (KEK), 
Tsukuba, Ibaraki 305-0801, Japan}
\affiliation{Department of Particle and Nuclear Physics
The Graduate University for Advanced Studies
Tsukuba Ibaraki 305-0801 Japan}
\author{Satoshi Nagaoka}
\email{nagaoka@post.kek.jp}
\affiliation{High Energy Accelerator Research Organization (KEK),
Tsukuba, Ibaraki 305-0801, Japan}

\begin{abstract}
We construct 
Green-Schwarz (GS) light-cone closed superstring theory from type IIB matrix
model. A GS light-cone string action is derived 
from two dimensional ${\cal N}=8$ $ U(n)$
noncommutative
 Yang-Mills (NCYM) by identifying noncommutative scale with string scale.
Supersymmetry transformation for the light-cone gauge action is also
 derived from supersymmetry transformation for IIB matrix model.
By identifying the physical states and interaction vertices, 
string theory is perturbatively reproduced.
\end{abstract}

\maketitle

\section{Introduction}

While the theory of general relativity classically 
gives a proper description of spacetime,
it brings unrenormalizable divergences by quantum effects.
In order to avoid the divergence of quantum gravity, 
we need to modify the geometry at short length scale.
Technically, we need to  
introduce an effective `cut-off' in the theory. 
Superstring theory naturally gives such a cut-off, since 
the ultra-violet divergence is removed by the cut-off 
at string scale $\alpha'$.
While string theory is well-defined only in ten dimensions, 
the structure of space-time, which includes 
the number of effectively emergent space-time dimensions,
should be decided by itself.
Type IIB matrix model is proposed as a nonperturbative formulation of
superstring theory \cite{IKKT}. Originally, it is derived by a matrix
regularization of Green-Schwarz type IIB superstring.
Light-cone superstring field theory of type IIB superstring 
is reproduced from a continuum limit of loop equations
for Wilson loops
in the large $N$ limit \cite{FKKT}.
Since this model is in some sense 
a Lorentz covariantly regularized theory,
not only space, but also even time may naturally 
emerge in this model.

In string theory, the effect of cut-off 
is seen through the T-duality transformation \cite{KY,SS}.
In a compactified space with the radius $R$, 
a string whose compactification scale $R$
is equivalent to that of $\alpha'/R$.
In type IIB matrix model, a similar kind of equivalence
is also seen. 
Although there is no dimensionful parameter in IIB matrix model,
once we select a certain background, then, the 
characteristic dimensionful parameter, which is called the noncommutative
parameter, emerges.
It is proposed that 
the self-dual scale of this theory exists at a noncommutative scale $\theta$
\cite{IIKK2,KTT,KN2}.
Another important property of type IIB matrix model is
supersymmetries.
In string theory, interaction vertex for GS light-cone 
closed superstring is completely determined by the supersymmetry
transformation.
In IIB matrix model,
the coupling to supergravity multiplet is determined by the
supersymmetry transformation \cite{vertex}.
The action of IIB matrix model can be derived by the 
dimensional reduction of ten dimensional super Yang-Mills (SYM) theory 
to zero dimension.
Each SYM with different dimensionality shows an interesting 
behavior and they are related with each other
in a large variety of way.
 In particular, two dimensional ${\cal N}=8$
$SU(N)$ SYM is proposed as a type IIA multi-string theory \cite{DVV}.
It is conjectured as an another candidate of nonperturbative
formulation of superstring theory.

In this paper, we search for the description of perturbative 
superstring theory from type IIB matrix model.
Type IIB matrix model naturally includes ${\cal N}=8$
 noncommutative 
Yang-Mills theory (NCYM) \cite{CDS,AIIKKT,Li}. In the commutative limit, 
it reduces to ordinary gauge theory.
In such a sense, type IIB matrix model contains
matrix string theory in the large $N$ limit.
In order to clarify and develop such a perspective,
we derive a Green-Schwarz (GS) superstring action
from type IIB matrix model.
We can formulate a string perturbation theory
and calculate multi-point correlation functions from $U(n)$ NCYM 
since multiple string worldsheets are derived from 
$U(n)$ NCYM.
Furthermore, we derive supersymmetry transformation for the GS light-cone
superstring action from type IIB matrix model.
The massless particle 
vertex operators for GS light-cone superstring theory are
determined by the requirement that they transform into one another under
supersymmetry transformations.
Since we reproduce the supersymmetry transformation for 
light-cone GS superstring from type IIB matrix model,
we can construct the corresponding 
vertex operators in the same procedure.

\section{Superstring from type IIB matrix model\label{2}}

\subsection{Superstring action}

Type IIB matrix model is defined by the action as
\begin{eqnarray}
S=-\frac{1}{g^2} \Tr
 \left(\frac{1}{4}[A^\mu,A^\nu][A_\mu,A_\nu]+\frac{1}{2}\bar{\psi} \Gamma^\mu
 [A_\mu,\psi] \right) \ , \label{action1}
\end{eqnarray}
where $\psi$ is a ten dimensional Majorana-Weyl spinor. $A_\mu 
(\mu=0,1,\cdots,9)$ and
$\psi$ are $N\times N$ Hermitian matrices. 
The coupling constant $g$ can be absorbed by the field redefinition
\begin{eqnarray}
A_\mu &\to& g^{1/2} A_\mu \ , \n
\psi  &\to& g^{3/4} \psi \ .
\end{eqnarray}
$d$ dimensional Euclidean NCYM
are obtained by expanding 
this action around $d$ dimensional flat background:
\begin{eqnarray} \label{theta}
[p_\mu,p_\nu]=i\theta_{\mu\nu} \ .
\end{eqnarray}
In this construction, two dimensional NCYM with
${\cal N}=8$ supersymmetry
is the lowest dimensional 
theory and the lagrangian is
written as
\begin{eqnarray}
S=-\frac{\theta}{8 \pi g^2} \int d^2 x  \tr \left(
[D^{\tilde{\mu}},D^{\tilde{\nu}}][D_{\tilde{\mu}},D_{\tilde{\nu}}] +2
[D^{\tilde{\mu}},\phi^i][D_{\tilde{\mu}},\phi_i] \right. \n 
+[\phi_i,\phi_j] 
[\phi_i,\phi_j]
\left. +2 \bar{\psi} \Gamma^{\tilde{\mu}} [D_{\tilde{\mu}},\psi]
+2 \bar{\psi} \Gamma_i 
[\phi_i,\psi]
\right)_* \ , \n
\label{2dNCYM}
\end{eqnarray}
where $\tilde{\mu},\tilde{\nu}=0,1$ and $i,j=2,\cdots ,9$.
$*$ product is described by 
\begin{eqnarray}
a * b =\exp \left( \frac{iC^{\mu\nu}}{2} \frac{\p^2}{ \p \xi^\mu \p
		\eta^\nu} \right) a(x+\xi) b(x+\eta) |_{\xi=\eta=0} \ .
\end{eqnarray}
$D_{\tilde{\mu}}$ are covariant derivative 
operators which contain the gauge
field as
\begin{eqnarray}
[D_{\tilde{\mu}}, \hat{o}]=[\hat{p}_{\tilde{\mu}}+\hat{a}_{\tilde{\mu}},
\hat{o}] \to
\frac{1}{i} \p_{\tilde{\mu}} o +a_{\tilde{\mu}} * o - o * 
a_{\tilde{\mu}} (x) \ .
\end{eqnarray}
Trace of the matrices maps into the integral of the functions as
\begin{eqnarray} \label{map}
\Tr \to \frac{\theta}{2 \pi} \tr \int d^2 x  \ .
\end{eqnarray}
Remaining trace in (\ref{2dNCYM})
is the trace over $U(n)$ gauge group in two dimension. 
Thus, two dimensional theory is constructed from type 
IIB matrix model. 

We will identify the perturbative string spectrum 
in the IR limit of two dimensional NCYM.
First of all, 

i) $*$ product goes to ordinary commutative product
since higher derivatives in the product can be neglected.
The action (\ref{2dNCYM}) becomes the commutative ${\cal N}=8$
$U(n)$ super Yang-Mills in this limit
\begin{eqnarray}
S&=&-\frac{\theta}{8 \pi g^2} \int d^2 x \tr \left(
F^2_{\tilde{\mu}\tilde{\nu}} +2 (D_{\tilde{\mu}} \phi_i)^2
+[\phi_i,\phi_j][\phi_i,\phi_j] 
\right. \n &&\left.
+2 \bar{\psi} \Gamma^{\tilde{\mu}}
D_{\tilde{\mu}} \psi +2 \bar{\psi} \Gamma_i [\phi_i,\psi] \right) \ .
\label{2dYM}
\end{eqnarray}
This action includes 8 matrix scalar fields $\phi_i$ 
and 16 matrix spinor fields 
$\psi= (s^a, s^{\dot{a}})$. These fields transform 
in $8_v$, $8_c$ and $8_s$ representations of SO(8) group.
The perturbative vacua of this action are represented by
the diagonal matrices $\phi_i =(\phi_{\text{diag}})_i$, which 
form the moduli space of this theory.

By assuming that all the eigenvalues of matrices do not coincide 
with each other at any points on the worldsheet,
all the excitations 
of off-diagonal modes become massive.
Then, 

ii) only diagonal elements are relevant
in the low energy limit
since massless excitations come from diagonal elements.
The contribution of the 
terms $[\phi_i,\phi_j][\phi_i,\phi_j]$ and $2 \bar{\psi} \Gamma_i
[\phi_i,\psi]$ vanish since diagonal terms commute. 
Gauge fields on two dimension, which come from the first term in 
(\ref{2dYM}), decouple from other fields.

In order to identify the action (\ref{2dYM}) 
with a light-cone superstring action,
we map the worldsheet coordinate system from $R^2$
coordinate into $R^1 \times S^1$ coordinate as
\begin{eqnarray}
z \equiv x_0+ix_1=e^{\tau+i\sigma} \ .
\end{eqnarray}
The origin of the $x$ coordinate is the special point where 
the vertex operators are inserted.
By the rescaling,
\begin{eqnarray}
\psi_R \to \frac{1}{\sqrt{z}} \psi_R \ , \quad 
\psi_L \to \frac{1}{\sqrt{\bar{z}}} \psi_L \ ,
\end{eqnarray}
we obtain the action for a single string with the winding number $w$
\begin{eqnarray}
S=-\frac{\theta}{4 \pi g^2}
\int_{0}^{\infty} d\tau \int_0^{2\pi w} d\sigma 
\left((\p_{\tau} \phi_i)^2 +(\p_\sigma \phi_i)^2
 \right. \n \left. 
+\bar{\psi} (\Gamma^+ \p_+ + \Gamma^- \p_- ) \psi
\right) \ . \label{GSaction2}
\end{eqnarray}
Multiple strings are obtained in general.
Thus, two dimensional NCYM (\ref{2dNCYM}) 
reduces to the GS light-cone superstring theory
in the IR limit.
A string with the winding number $w$ along the $\sigma$ direction is
described in Figure 1.
\begin{figure}[bht]
\begin{center}
\includegraphics[height=3cm]{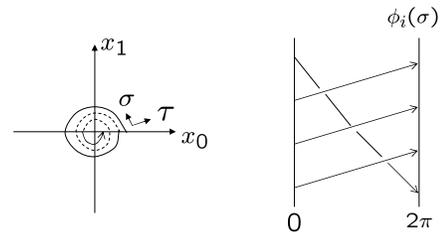} 
\caption{A string winds $w$ times along $\sigma$ direction.}
\end{center}
\end{figure}
We have identified $\phi_i (\sigma)=\phi_i (\sigma+ 2 \pi w)$
in this case.
$\theta$ is introduced as a characteristic scale of background (\ref{theta}).
In this background, $\theta$ is a unique dimensionful parameter.
GS light-cone superstring
action is obtained by identifying
$\frac{\theta}{4\pi g^2}\equiv
\frac{1}{4 \pi \alpha'}$. Since (\ref{GSaction2}) is obtained by
taking the commutative limit,
the winding number $w$ has to be large.
We can reproduce the Veneziano type formula 
if we 
sum up the contribution from infinite towers of the spectrum.
On-shell condition is relevant only to external modes.
In fact, we calculate four point amplitude from type IIB matrix model 
where stringy effect is relevant \cite{KN4}.

The winding number $w$ along the angular direction can be reinterpreted
as light-cone 
momenta $p^+$ in a T-dual interpretation.
In this interpretation, we obtain type IIA superstring theory.
Indeed, vertex operators in GS light-cone superstring
for type IIA supergravity multiplet are
constructed \cite{KN4} 
from original vertex 
operators for type IIB matrix model \cite{vertex,ITU,KMS}.
Vertex operators in matrix model are utilized to show the localization of
gravity on D-brane \cite{KN1} like a Randall-Sundrum model \cite{RS}.

The duality relation is shown in Figure 2. 
We show our construction along with 
 Dijkgraaf-Verlinde-Verlinde's matrix string theory.
Two dimensional NCYM obtained here is the action of a D-string.
In DVV's matrix string theory, two dimensional SYM action,
which is obtained 
by the compactification along 9-direction 
(T-duality transformation) from BFSS matrix model, is the action of D-strings. 
The former is uncompactified theory, although the latter is compactified
along 9-direction, which is the direction longitudinal to the
worldvolume of the D-string. They obtain type IIA string in a
T-dual interpretation along the light-cone direction.
On the other hand, we obtain type IIA string 
by radial quantization.
\begin{figure}[hbt]
\begin{center}
\includegraphics[height=4cm]{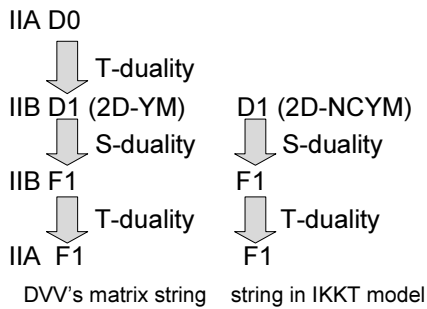} 
\caption{Duality relation between type IIA and type IIB string is shown.}
\end{center}
\end{figure}

\subsection{Supersymmetry transformation}

${\cal N}=2$ supersymmetry transformation in IIB matrix model
is written as
\begin{eqnarray}
\delta^{(1)} \psi &=&\frac{i}{2} [A_\mu,A_\nu]\Gamma^{\mu\nu} 
\epsilon \ , \n
\delta^{(1)} A_\mu &=&i \bar{\epsilon} \Gamma_\mu \psi \ , \n
\delta^{(2)}\psi &=&-\eta \ , \n
\delta^{(2)} A_\mu &=&0 \ .
\end{eqnarray}
In the two dimensional NC background,
this transformation can be written as 8 scalar fields, 16 spinor
fields and 2 gauge fields.
In the low energy limit, it
reduces to 
\begin{eqnarray}
\delta^{(1)} s_a &=&- \dot{\phi}^i \gamma_{a\dot{a}}^i
\epsilon^{\dot{a}} \ , \quad
\delta^{(1)} s_{\dot{a}} =
- \tilde{\dot{\phi}}^i \gamma_{\dot{a}a}^i
\epsilon^{a}
\ ,
\n
\delta^{(1)} \phi_i &=&2
(\bar{\epsilon}^{\dot{a}} \gamma_{a\dot{a}}^i s^a+
\bar{\epsilon^{a}} \gamma_{\dot{a}a}^i s^{\dot{a}})
\ , \n
\delta^{(2)}s_a &=&- \eta^{a} \ , 
\quad
\delta^{(2)}s_{\dot{a}} =- \eta^{\dot{a}} \ , 
\n
\delta^{(2)} \phi_i &=&0 \ ,
\end{eqnarray}
where we have redefined $\eta^a \to \eta^a+\theta \epsilon$,
$\eta^{\dot{a}}\to \eta^{\dot{a}}-\theta \epsilon$ to 
absorb the constant shift.
$\gamma^i_{a\dot{a}}$ are Clebsch-Gordan coefficients for 
coupling three inequivalent SO(8) representations.
Indeed, these transformations leave the Green-Schwarz light-cone 
string action (\ref{GSaction2}) invariant.

The sixteen supersymmetry charges are $8_s$ and $8_c$ given by
\begin{eqnarray}
Q^a&=& s_0^a \ , \n 
Q^{\dot{a}}&=& \sqrt{2} \gamma_{a\dot{a}}^i \sum_{-\infty}^{\infty}
s_{-n}^a \alpha_n^i \ ,
\end{eqnarray}
for the left mover and 
\begin{eqnarray}
\tilde{Q}^a&=& s_0^{\dot{a}} \ , \n 
\tilde{Q}^{\dot{a}}&=& \sqrt{2} \gamma_{\dot{a}a}^i \sum_{-\infty}^{\infty}
s_{-n}^{\dot{a}} \tilde{\alpha}_n^i \ ,
\end{eqnarray}
for the right mover 
where quantization of superstring action involves 4 sets of modes
denoted by $\alpha_n^i$, $\tilde{\alpha}_n^i$, 
$s_n^a$ and $s_n^{\dot{a}}$. They satisfy the relations 
\begin{eqnarray}
[\alpha_m^i,\alpha_n^j]&=&m \delta_{m+n} \delta^{ij} \ , \n \
[\tilde{\alpha}_m^i,\tilde{\alpha}_n^j]&=&m \delta_{m+n} \delta^{ij} \ , \n
\{s_m^a, s_n^b \}&=& \delta_{m+n}\delta^{ab} \ , \n
\{s_m^{\dot{a}}, s_n^{\dot{b}} \}&=& \delta_{m+n}\delta^{\dot{a}\dot{b}} 
\ .
\end{eqnarray}
Let us consider the left moving bosonic (vector) and
fermionic (spinor) vertex operators
\begin{eqnarray}
V_B (\zeta,k)&=&\zeta \cdot B e^{i k \cdot \phi} \ , \n
V_F(u,k)&=&uFe^{ik\cdot \phi} \ ,
\end{eqnarray}
where $\zeta$ is a polarization vector which represents 
the wave function for the vector state and $u$ represents the
wave function for the spinor states.
The coefficients $B$ and $F$ are determined by 
the supersymmetry transformation
\begin{eqnarray}
[\eta^a Q^a,V_F (u,k)] &\approx& V_B (\tilde{\zeta},k) \ , \quad \n \
[\eta^a Q^a,V_B (u,k)] &\approx& V_F (\tilde{u},k) \ .
\end{eqnarray}
The symbol $\approx$ means that equality is only required for on-shell matrix
elements. The structure of the vertex operators is uniquely determined by
the requirement of global supersymmetry.
The closed string vertex operators are constructed by the product of 
open string vertex operators
\begin{eqnarray}
V(\sigma,\tau)=V_R (\tau-\sigma) V_L (\tau+\sigma) \ ,
\end{eqnarray}
where the left moving operator $V_L(\tau-\sigma)$ and the 
right moving operator 
$V_R (\tau+\sigma)$ 
are either bosonic $V_B$ or fermionic $V_F$.
Explicit forms of vertex operators are constructed \cite{KN4} within the 
kinematics where external momentum along the light-cone direction is
$k^+=0$, since $A^-$ has serious ordering problem in the light-cone
gauge \footnote{In order to resolve this problem, a separate Fock space 
for each string is introduced.}.

\section{Conclusion \label{6}}

We have derived Green-Schwarz light-cone superstring theory
from type IIB matrix model.
As illustrating in Figure 3, (1) two dimensional background in type 
IIB matrix model reduces to Green-Schwarz superstring action 
in the low energy limit.
The field $\phi$ is interpreted as a displacement of the
{\it type IIA superstring}, since this theory is derived from type
IIB matrix model in a {\it T-dual interpretation}.
This is similar to the derivation of matrix string by Dijkgraaf, Verlinde and
Verlinde.
(2) In order to derive string perturbation from type IIB matrix model,
we have directly derived supersymmetry transformation for GS light-cone 
string from type IIB matrix model.
The supersymmetry transformation also shows symmetries of the type IIA 
superstring theory.
Since vertex operators are
determined by the requirement that they transform into one another under
supersymmetry transformations, we can uniquely determine the vertex operators
to calculate the amplitude of
multi-point function.
On the other hand,
the vertex operators in type IIB matrix model are constructed in 
\cite{vertex,ITU,KMS} where external momenta are carried by the Wilson
lines.
(3) We can derive the vertex operators for GS type IIA 
light-cone string directly
from type IIB matrix model, which will be done in \cite{KN4}.

In this paper, as a first step to describe the string 
interaction in type IIB matrix model, 
we have obtained the second quantized worldsheet theory
by analyzing $U(n)$ NCYM.
Fundamental scale $\theta$ and the maximal supersymmetry play crucial
roles for the identification.
As we have identified the noncommutative scale with string scale, we need to
understand its significance more deeply such as the relation with
space-time uncertainty principle \cite{Yoneya}.
If the Wilson line carries the momentum $p^+$,
it extends
$ {\cal O} (\frac{|p^+|}{\theta}) \sim {\cal O} (\alpha' |p^+|)$
in the orthogonal direction.
Thus, the length of Wilson line may be identified with 
light-cone momentum, which is familiar in
the light-cone formulation of superstring theory.
In order to treat multi-string interactions, we need to construct 
some new types of formulations like closed superstring field theory. 
In Dijkgraaf, Verlinde and Verlinde's matrix string theory, 
they introduce an interaction of multi strings as a
recombination of intersecting (D-)strings \cite{HN}.
It may also be interesting to consider this effect within 
type IIB matrix model. 

\vspace*{1cm}
\begin{figure}[hbt]
\begin{center}
\includegraphics[height=2cm]{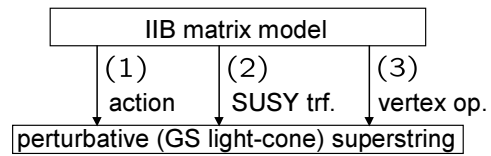} 
\caption{
Green-Schwarz light-cone superstring theory are perturbatively obtained 
by three ways. }
\end{center}
\end{figure}

\begin{acknowledgments}
This work is supported in part by the Grant-in-Aid for Scientific 
Research from the Ministry of Education, Science and Culture of
Japan.
The work of S.N. is supported in part by the Research Fellowship of
the Japan Society for the Promotion of Science for Young Scientists.
\end{acknowledgments}

\appendix


\begin{thebibliography}{99}

\bibitem{IKKT}
N. Ishibashi, H. Kawai, Y. Kitazawa and A. Tsuchiya,
Nucl. Phys. {\bf B498} (1997) 467, hep-th/9612115.
\bibitem{FKKT}
 M. Fukuma, H. Kawai, Y. Kitazawa and A. Tsuchiya,
Nucl. Phys. {\bf B510} (1998) 158,
hep-th/9705128.
\bibitem{KY}
K. Kikkawa and M. Yamasaki,
Phys. Lett. {\bf B149} (1984) 357.
\bibitem{SS}
N. Sakai and I. Senda,
Prog. Theor. Phys. {\bf 75} (1986) 692, [Erratum-ibid. {\bf 77} (1987)
	773]. 
\bibitem{IIKK2}
N. Ishibashi, S. Iso, H. Kawai and Y. Kitazawa,
Nucl. Phys. {\bf B583} (2000) 159,
hep-th/0004038.
\bibitem{KTT}
Y. Kitazawa, Y. Takayama and D. Tomino,
Nucl. Phys. {\bf B715} (2005) 665,
hep-th/0412312.
\bibitem{KN2}
Y. Kitazawa and S. Nagaoka,
Phys. Rev. {\bf D75} (2007) 046007,
hep-th/0611056.
\bibitem{vertex}
Y. Kitazawa,
JHEP {\bf 0204} (2002) 004,
hep-th/0201218.
\bibitem{DVV}
 R. Dijkgraaf, E. Verlinde and H. Verlinde,
Nucl. Phys. {\bf B500} (1997) 43,
hep-th/9703030.
\bibitem{CDS} A. Connes, M. Douglas and A. Schwarz,
JHEP{\bf 9802} (1998) 003, hep-th/9711162.
\bibitem{AIIKKT}
H. Aoki, N. Ishibashi, S. Iso, H. Kawai, Y. Kitazawa
and T. Tada,
Nucl. Phys. {\bf 565} (2000) 176,
hep-th/9908141.
\bibitem{Li}
M. Li, 
Nucl. Phys. {\bf B499} (1997) 149, hep-th/9612222.
\bibitem{KN4}
Y. Kitazawa and S. Nagaoka,
in progress.
\bibitem{ITU}
S. Iso, H. Terachi and H. Umetsu,
Phys. Rev. {\bf D70} (2004) 125005,
hep-th/0410182.
\bibitem{KMS}
 Y. Kitazawa, S. Mizoguchi and O. Saito,
Phys. Rev. {\bf D75} (2007) 106002,
hep-th/0612080.
\bibitem{KN1}
Y. Kitazawa and S. Nagaoka,
JHEP {\bf 0602} (2006) 001,
hep-th/0512204.
\bibitem{RS}
L. Randall and R. Sundrum,
Phys. Rev. Lett. {\bf 83} (1999) 4690,
hep-th/9906064.
\bibitem{Yoneya}
T. Yoneya,
Mod. Phys. Lett. {A4} (1989) 1587.
\bibitem{HN}
 K. Hashimoto and S. Nagaoka,
JHEP {\bf 0306} (2003) 034,
hep-th/0303204.

\end{thebibliography}
\end{document}